\documentclass[11pt, preprint]{aastex}

\shorttitle{Biases in mass estimates of dSph galaxies}

\shortauthors{{\L}okas et al.}

\begin{document}

\title{Biases in mass estimates of dSph galaxies}

\author{Ewa L. {\L}okas\altaffilmark{1}, Stelios Kazantzidis\altaffilmark{2}, Jaros{\l}aw Klimentowski\altaffilmark{1}
and Lucio Mayer\altaffilmark{3}}

\altaffiltext{1}{Nicolaus Copernicus Astronomical Center, 00-716 Warsaw, Poland; lokas@camk.edu.pl}
\altaffiltext{2}{Center for Cosmology and Astro-Particle Physics,
    The Ohio State University, Columbus, OH 43210, USA; stelios@mps.ohio-state.edu}
\altaffiltext{3}{Institute for Theoretical Physics, University of Z\"urich, CH-8057 Z\"urich, Switzerland;
        lucio@phys.ethz.ch}

\begin{abstract}
Using a high resolution $N$-body simulation of a two-component dwarf galaxy orbiting in the
potential of the Milky Way, we study two effects that lead to significant biases in mass
estimates of dwarf spheroidal galaxies. Both are due to the strong tidal interaction of initially disky dwarfs with
their host. The tidal stripping of dwarf stars leads to the formation of strong tidal tails that
are typically aligned with the line of sight of an observer positioned close to the host center.
The stars from the tails contaminate the kinematic samples leading to a velocity dispersion
profile increasing with the projected radius and resulting in an overestimate of mass. The tidal stirring of the
dwarf leads to the morphological transformation of the initial stellar disk into a bar and then a
spheroid. The distribution of stars in the dwarf remains non-spherical for a long time leading to
an overestimate of its mass if it is observed along the major axis and an underestimate if it seen
in the perpendicular direction.
\end{abstract}


\section{Introduction}

\begin{figure}[t]
\begin{center}
    \includegraphics[width=16cm]{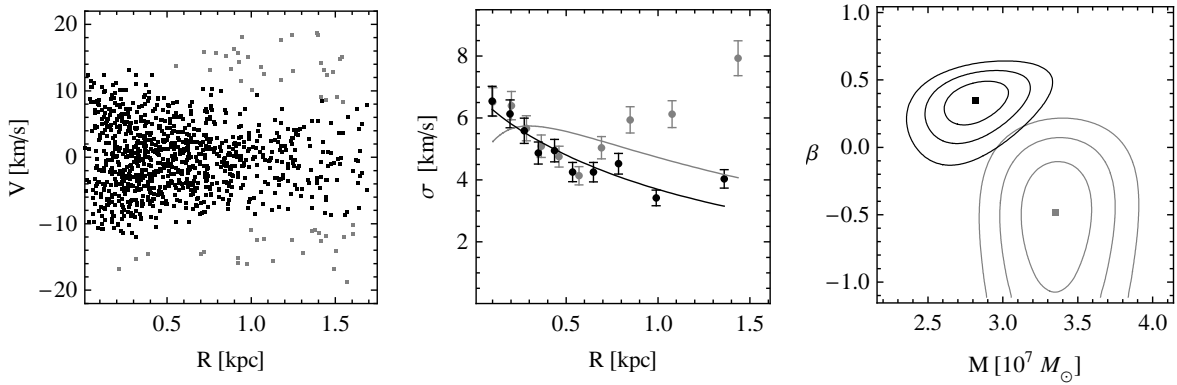}
\end{center}
\caption{An example of a kinematic sample of a thousand stars (left panel), the velocity dispersion profiles measured
from it (middle panel) and the contours showing the inferred mass and anisotropy (right panel). In each panel the
gray (black) color refers to the sample contaminated by (cleaned of) interloper stars from the tails.}
\label{fig1}
\end{figure}

For the purpose of this study we used a high-resolution simulation of a two-component dwarf galaxy orbiting
in the gravitational potential of the Milky Way (Klimentowski et al. 2007, 2009a). The dwarf
progenitor of total mass $M = 4.3 \times 10^9 M_{\odot}$ consisted of a baryonic disk embedded in a dark matter halo.
The initial mass of the disk was $1.5 \times 10^8 M_{\odot}$ and the mass of the NFW dark matter halo was $4.1 \times
10^9 M_{\odot}$. The dwarf galaxy was placed on an eccentric orbit around the host galaxy with apocenter to
pericenter ratio of $r_{\rm a}/r_{\rm p}=120/25$ kpc and the disk initially inclined by 45$^\circ$ to the
orbital plane. The evolution was followed for 10 Gyr corresponding to five orbital times. The host galaxy was modelled
by a static gravitational potential assumed to have the present-day properties of the Milky Way as described by mass
model A1 of Klypin et al. (2002).

During the evolution the dwarf galaxy is strongly affected by the tidal field of the host which results in a
significant mass loss, the morphological transformation from a disk to a bar and then a spheroid and the transition
from the streaming to the random motion of the stars. At all times, including the final stage, the core of the dwarf
galaxy remains gravitationally bound but is surrounded by pronounced tidal tails and its shape departs from spherical.
If dwarf spheroidal galaxies of the Local Group indeed formed as envisioned by this so-called tidal stirring scenario
(Mayer et al. 2001, 2007), these effects must be taken into account when modeling their masses. Here we provide quantitative
estimates of the biases imposed by the contamination of kinematic samples by unbound stars from the tails and by the
departures from sphericity of the stellar component.

\section{Contamination of kinematic samples by tidally stripped stars}

The left panel of Fig.~\ref{fig1} shows a kinematic sample of a thousand stars randomly selected from the final output
of the simulation (where the dwarf galaxy is at apocenter). This mock data set shows line-of-sight velocities as a
function of the projected distance $R$ as would be measured by a distant observer. The observation is made along the
tidal tails, which is a typical orientation of the debris for an observer placed near the center of the Milky Way
(Klimentowski et al. 2009b).
Only the data within the standard velocity cut-off of $\pm 3\sigma_0$ (where $\sigma_0$ is the central
velocity dispersion of the stars) from the systemic velocity of the dwarf are shown. From this data set we obtain the
velocity dispersion profile shown in the middle panel of Fig.~\ref{fig1} with gray data points with sampling errors.
We then fit to this profile the solutions of the Jeans equation for spherical systems, assuming that mass
traces light (the distribution of stars is measured for the same line of sight), and adjusting two parameters: the total
mass $M$ and the anisotropy parameter of stellar orbits $\beta=1-\sigma_\theta^2/\sigma_r^2$. The best-fitting solution
is shown as a gray line in the middle panel of Fig.~\ref{fig1} and the constraints on $M$ and $\beta$ are shown with
the gray $1\sigma,2\sigma$ and $3\sigma$ contours in the right panel of Fig.~\ref{fig1}.

The increase of the velocity dispersion profile for $R>0.7$ kpc is due to the contamination of the sample by
unbound stars from the tidal tails. Note that the true values of the mass and anisotropy of this dwarf galaxy within a
sphere of $r<1.7$ kpc measured in the simulation are respectively $M=2.1 \times 10^7 M_{\odot}$ and $\beta=0.33$
({\L}okas et al. 2010).
The use of the contaminated sample thus leads to an overestimate of the mass and an underestimate of the
anisotropy (tangential rather than mildly radial orbits are inferred). The amount of this bias will depend on the
actual orientation of the tails and the initial velocity cut-off applied to the sample, but the presented example shows
that it can be quite large. Since the fit is rather bad one may try to improve it by introducing a more extended
dark matter halo. The tidal stirring scenario predicts, however, that tidally affected dwarfs do not possess such halos
and their mass approximately traces light (Mayer et al. 2001, {\L}okas et al. 2010).

\begin{figure}[t]
\begin{center}
    \includegraphics[width=16cm]{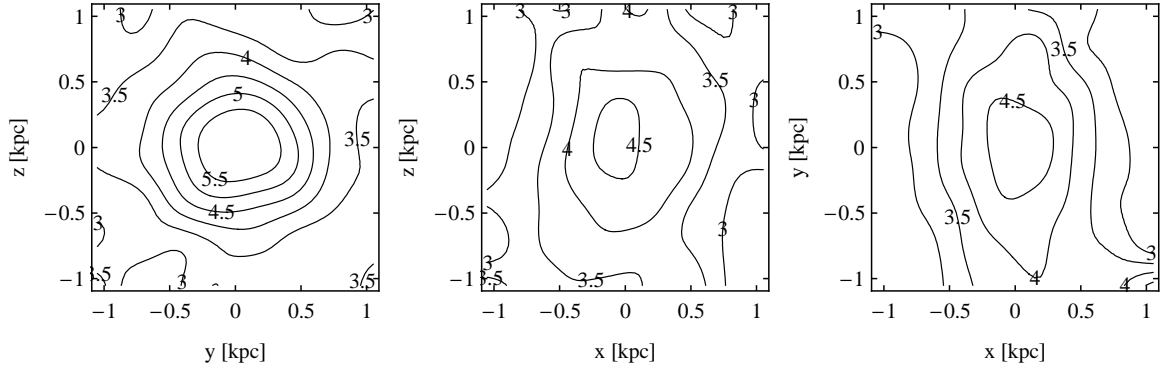}
\end{center}
  \caption{Maps of the line-of-sight velocity dispersion of the stars measured along the long (left panel) and the two
short axes (middle and right panel). Numbers give the values of the velocity dispersion in km s$^{-1}$.}
\label{fig2}
\end{figure}

The problem is solved when the kinematic sample is cleaned of interlopers (Klimentowski et al. 2007, {\L}okas et al.
2008, {\L}okas 2009). The stars marked
by the gray dots in the left panel of Fig.~\ref{fig1} have been removed by a scheme based on assigning a maximum
velocity allowed for a star at a given $R$ which was extensively tested on simulated data (Klimentowski et al. 2007). The
resulting velocity dispersion profile (black points in the middle panel of Fig.~\ref{fig1}) is then no longer increasing
with radius and our simple two-parameter model fits the data much better. The best-fitting mass and
anisotropy (black contours in the right panel of Fig.~\ref{fig1}) obtained in this case are in very good agreement with
the real values.

\section{The effect of non-sphericity of the dwarf}

\begin{figure}[t]
\begin{center}
    \includegraphics[width=16cm]{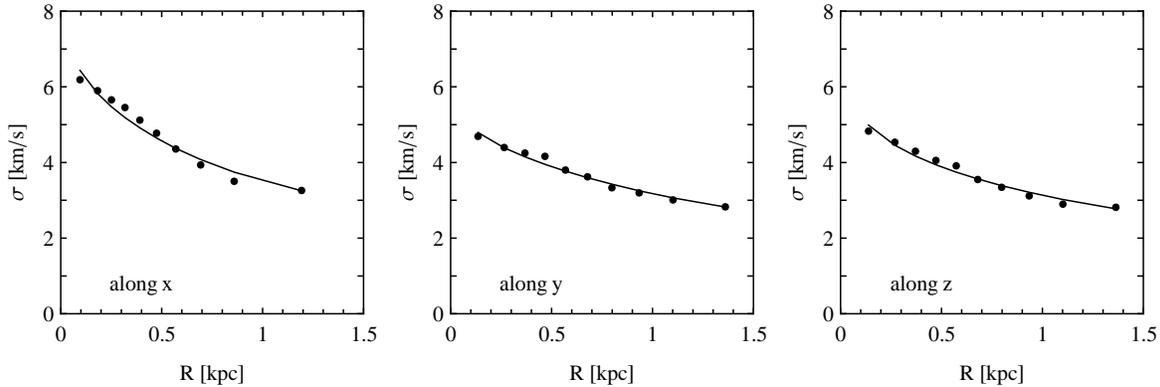}
\end{center}
  \caption{Line-of-sight velocity dispersion profiles of the stars when the dwarf is observed along the long (left
	panel) and one of the two short axes (middle and right panel). The solid lines show the best-fitting solutions
	of the Jeans equation.}
\label{fig3}
\end{figure}

The products of tidal stirring are typically non-spherical. In the example discussed here the distribution of the stars
in the final stage is a prolate spheroid with the short-to-long axis ratio of 0.6 (Klimentowski et al. 2009a).
The shape affects the
velocity distribution in the dwarf which retains some of the radial orbits characteristic of the long-lasting bar-like
phase of the evolution where the axis ratio was much smaller. Fig.~\ref{fig2} shows the maps of the line-of-sight
velocity dispersion of the stars seen along the long ($x$) and the two short ($y$ and $z$) axes (from the left to the
right panel). The dispersion is larger when viewed along the $x$ axis and it decreases more steeply with radius.

The same effect is seen in Fig.~\ref{fig3} where we plot the velocity dispersion profiles measured in a traditional
way, in concentric shells of projected radius $R$: when measured along the long axis of the stellar component the
dispersion reaches higher values in the center and decreases more steeply with $R$. Note that here only the stars
within the sphere of radius $r<1.7$ kpc were used so no contamination from the tidal tails is present and all profiles
decrease with radius. In addition, we used all (i.e. almost 38 thousand) stars, rather than a sample of a thousand to
avoid the noise due to sampling errors (those are then too small to be visible in Fig.~\ref{fig3} and are not shown).

As in the previous section, we fitted the dispersion profiles with the solutions of the Jeans equation for spherical
systems, adjusting the total mass and anisotropy, assuming that mass follows light (the stellar density profile was
fitted separately for every line of sight). For the view along $x$ the mass within 1.7 kpc is overestimated by 7\%, for
the view along $y$ and $z$ it is underestimated by 17-18\%. Interestingly, the anisotropy is
recovered almost perfectly, with the best-fitting values within the range of $\beta=0.2$-0.3 in all cases ({\L}okas et
al. 2010).

\acknowledgments

This research was partially supported by the
Polish Ministry of Science and Higher Education
under grant NN203025333.


\begin{thebibliography}{9}

\bibitem[{Klimentowski et al.}(2007)]{K2007}
J.~Klimentowski, E.~L.~{\L}okas, S.~Kazantzidis, F.~Prada, L.~Mayer, G.~A.~Mamon,
\emph{MNRAS} \textbf{378}, 353--368 (2007).

\bibitem[{Klimentowski et al.}(2009a)]{K2009a}
J.~Klimentowski, E.~L.~{\L}okas, S.~Kazantzidis, L.~Mayer, G.~A.~Mamon,
\emph{MNRAS} \textbf{397}, 2015--2029 (2009).

\bibitem[{Klypin et al.}(2002)]{KZS2002}
A.~Klypin, H.~Zhao, R.~S.~Somerville, \emph{ApJ} \textbf{573}, 597--613 (2002).

\bibitem[{Mayer et al.}(2001)]{M2001}
L.~Mayer, F.~Governato, M.~Colpi, B.~Moore, T.~Quinn, J.~Wadsley, J.~Stadel, G.~Lake,
\emph{ApJ} \textbf{559}, 754--784 (2001).

\bibitem[{Mayer et al.}(2007)]{M2007}
L.~Mayer, S.~Kazantzidis, C.~Mastropietro, J.~Wadsley,
	\emph{Nature} \textbf{445}, 738--740 (2007).

\bibitem[{Klimentowski et al.}(2009b)]{K2009b}
J.~Klimentowski, E.~L.~{\L}okas, S.~Kazantzidis, L.~Mayer, G.~A.~Mamon, F.~Prada,
\emph{MNRAS} \textbf{400}, 2162--2168 (2009).

\bibitem[{{\L}okas et al.}(2010)]{L2010}
E.~L.~{\L}okas, S.~Kazantzidis, J.~Klimentowski, L.~Mayer, S.~Callegari,
\emph{ApJ} \textbf{708}, 1032--1047 (2010).

\bibitem[{{\L}okas et al.}(2008)]{L2008}
E.~L.~{\L}okas, J.~Klimentowski, S.~Kazantzidis, L.~Mayer,
\emph{MNRAS} \textbf{390}, 625--634 (2008).

\bibitem[{{\L}okas}(2009)]{L2009}
E.~L.~{\L}okas,
\emph{MNRAS} \textbf{394}, L102--L106 (2009).



\end{thebibliography}
\end{document}